\documentclass[11pt]{article}
\usepackage{amsmath,amssymb,color,cite}
\usepackage{graphicx}
\usepackage{epstopdf}
\usepackage{geometry}
\numberwithin{equation} {section}
\begin{document}
\title{{\bf Water in Carbon Nanotubes Is Not the Same Old Stuff} \\ \vspace{1 cm} \begin{small}G. Reiter$^1$,C.  Burnham$^{1}$,D. Homouz$^1$,P.M. Platzman$^2$, J. Mayers$^3$, T. Abdul-Redah $^3$,A. P. Moravsky$^4$,J.C. Li$^5$, C-K Loong$^6$,A.I.Kolesnikov $^6 $ \\
\vspace{.5 cm}
$^1$ Physics Department, University of Houston, 4800 Calhoun Road,
Houston, Texas 77204, USA \\
$^2$ Bell Labs, 600 Mountain Ave, Murray Hill, New Jersey, 07974,USA\\
$^3$ ISIS Facility, Rutherford Appleton Laboratory, Chilton, Didcot, Oxfordshire,
OX11 0QX, UK\\
$^4$MER Corporation, 7960 S. Kolb Road, Tucson, Arizona 85706, USA\\
$^5$Department of Physics, UMIST, PO Box 88, Manchester, M60, 1QD, UK\\
$^6$ Intense Pulsed Neutron Source, Argonne National Laboratory,S. Cass Avenue, Argonne Illinois 60439 USA \\
 \end{small}}

\author{}

\date{\today}
\maketitle
\begin{abstract}
The momentum distribution of the protons in ice Ih, ice VI, high density amorphous ice and water in carbon nanotubes at low temperatures has been measured using deep inelastic neutron scattering. We find that the momentum distribution for the water in the nanotubes is qualitatively unlike that in any other phase of water or ice. The kinetic energy of the protons is 35mev less than that in ice Ih at the same temperature, and the high momentum tail of the distribution, characteristic of the molecular covalent bond and the stretch mode of the proton in the hydrogen bonds, is not present. We observe a phase transition between 230K and 268K in the nanotube data. The high momentum tail  is present  in the higher temperature measurement, which resembles that of ice Ih at the same temperature.   Molecular dynamics simulations show the phase transition to be associated with the reordering  of the hydrogen bonds of the 2-D ice layer, coating the interior of the nanotube at low temperatures, into a  3-D  structure at 268K. We conclude that the protons in the hydrogen bonds in the 2-D ice layer are coherently delocalized, and  that the 2-D ice layer is a qualitatively new phase of ice. 
\end{abstract}

Water in carbon nanotubes is of interest as a model system for the study of water in quasi one-dimensional confined spaces\cite{hum}, where otherwise inaccessible liquid and glassy phases exist over a wide range of temperatures.  It has posssible applications to nanotechnology\cite{aj, mey} and the understanding of transport in biological pores.\cite{no,ka, maj } There is a consensus, arising from simulations of the structure,  that for sufficiently large nanotubes, the water entering initially forms an ordered 2-D square ice layer lining the nanotube. \cite{kog, no, ref1}. The momentum distribution of the protons, which, at the temperature of the measurements, is due almost entirely to zero point motion,  is a direct reflection of the structure of their local environment. It  can be measured by Deep Inelastic Neutron Scattering(DINS), also called Neutron Compton Scattering.\cite{rs,rmn, rmp,braz} . We find from these measurements, that the protons in the  ice layer are in a unique quantum state, qualitatively different from that of protons in the other phases of ice that we have measured, ice Ih, ice VI, high density amorphous ice(hda). The kinetic energy of the protons is 35 mev less than that of the protons in ice Ih at the same temperature, and the high momentum tail, characteristic  of the covalent bond in the water molecule, is missing.  There is a transition between 230K and 268K to a 3-D coordinated state that resembles the other phases of ice in the value of its kinetic energy and the presence of a high momentum tail in the momentum distribution. This is above the value (200K) predicted by our simulations of the structure, which are, however, classical, and do not satisfactorily include the quantum effects discussed here. 

 DINS  is inelastic neutron scattering in the limit of large momentum transfer, $\vec{q}$(30-100 $\AA^{-1})$. In this limit,  the neutrons scatter from the individual protons in the same manner that freely moving  particles scatter from each other. The fraction of neutrons scattered into a given angle with a given energy depends  only on the probability that the proton had a particular momentum  at the time it was struck by the neutron,  n($\vec{p}$). There is scattering of the neutrons off of the other ions as well, with the center of the peak due to an ion of mass M located at an energy of $\frac{\hbar^{2}q^2}{2M}$  Due to the much heavier mass of the carbon and oxygen, this scattering is easily separated from that of the protons.   We  actually measure directly the usual neutron scattering function, $S(\vec{q}, \omega)$, which in the limit of large momentum transfer has the impulse approximation form\cite{rs}
 
\begin{equation}
\label{defia15}
S_{IA}({\bf q},\omega)= \int n(\vec{p})\delta(\omega-{\frac{\hbar q^2}{2M}}-\vec{p}\cdot\vec{q})d\vec{p}
\end{equation}

The momentum distribution, $n(\vec{p})$ can be extracted from the measurements in a manner described in detail in earlier work\cite{rs,rmn,braz}. $S_{IA}({\bf q},\omega)$ is represented as a series expansion in Hermite polynomials. Small corrections due to deviations from the impulse approximation are added, the total  convolved with the instrumental resolution function, and the coefficients in the series expansion determined by a least squares fitting procedure. The measured  n(p) depends only on the magnitude of $\vec{p}$, since the nanotube sample is a randomly oriented powder.

$n(p)$ is then given by the expansion 
 \begin{equation}
 \label{inv}
 n(p) = {e^-{{p^2\over{2\sigma^2}}}\over{(\sqrt{2\pi}\sigma)^3}}
 \sum\limits_{n=0}^{\infty} a_{n}(-1)^n L_n^{{1\over
 2}}({{p^2\over{2\sigma^2}}})
\end{equation}
where the $ L_n^{{1\over
 2}}({{p^2\over{2\sigma^2}}})$ are associated Laguerre polynomials, and the $a_n$ are arbitrary coefficients to be determined by the least square fitting process. $a_0=1$. This expansion is complete, and can represent any function. We find that terms with n$>$5 are not statistically significant in these experiments. This definition of the coefficients  is chosen to be consistent with our earlier definition for the measurement of bulk ice and water.\cite{ braz} 
 
The parameter $\sigma$ that determines the width of the momentum distribution is related to the kinetic energy of the proton by K.E.=6.27$\sigma^2$ when $\sigma$ is expressed in inverse angstroms and the kinetic energy in meV,  independently  of any of the remaining coefficients if the term with n=1 is omitted in Eq. \ref{inv}.\cite{braz} The errors in the measured momentum distribution are related to the uncertainties in the coefficients in the expansion\cite{rmn}, which are obtained from the least squares fitting program, making a point by point calculation of the probable error possible.

The experiments were all done on the Vesuvio instrument at ISIS.  There were 28 detectors in all, arranged symmetrically around the beam.  Since the signal is isotropic, every detector gives potentially the same information. The simulations discussed were done as in Ref.\cite{ref1}. The 3 g sample of purified single walled carbon nanotubes (SWNT) was prepared as follows. The raw material, containing about  20 wt.$\%$ of SWNT, was obtained by direct current arc vaporization of graphite/metal composite  rods. The metal component consisted of Co/Ni catalyst in a 3:1 mixture. The purification of  SWNT was achieved by leaching out the metal catalyst with hydrochloric acid followed by  oxidation of non-tube carbon components by air at 300-600$^o$C, similar to the known procedure\cite{ch}. The purification procedure yields totally opened nanotubes  $14\pm 1\AA$ diameter and  about 10 $\mu$m in length as revealed by electron microscopy observation using HRTEM and SEM.    To fill nanotubes with water, the dry SWNT sample was first exposed in water vapor at 110¡C for  2h in an enclosed environment. The excess water adsorbed in the exterior of the nanotubes was  then extracted by evaporation at 45¡C to the final H2O/SWNT mass ratio of about 11.5 wt.$\%$.     High-pressure phase ice-VI and hda-ice were prepared  from double  distilled H2O water by pressurising the initial hexagonal ice-Ih in a piston-cylinder pressure cell  to 15 kbar, at about 270 K temperature for ice-VI and at 77 K for hda-ice. The ice-VI sample was  then cooled to 77 K at 15 kbar. The pressure was released and both samples were then recovered  at the low temperature (T=77 K).    

 In this isotropic case, 
\begin{equation}
\label{ysc}
 S_{IA}({\bf q},\omega)=\frac{ M}{q}J(y)=\frac{ M}{ q}  \int n(p)~\delta ( y-{\vec p}{\cdot \hat {q})\;d{\bf p}}
\end{equation}
where $y=\frac{M}{q}\left(  \omega - { {\hbar q^2}\over{2M}} \right)$ and $\hat{q}$ is a unit vector in the direction of the momentum transfer. 
We show in Fig.1a the signal for J(y) obtained directly from the data by adding  the signal that comes from intervals of time for each detector that map into an interval $\Delta y$, there being 100 such $\Delta y$ intervals in all. There is some systematic distortion  that arises from binning in this way, as  the final state corrections to the impulse approximation do not obey y scaling, and the binning procedure itself introduces some discreteness noise. The signal shown in Fig 1a includes the instrumental resolution. Nevertheless, it is clear to the eye that the 5K  data is considerably narrower than that at 268K. The proper way to fit the entire data set is with the polynomial expansion, as described above. The results for J(y) extracted in this way are shown in Fig 1b.  The parameters characterizing the fits are shown in Table 1, along with those for the other samples we have measured. The small statistical errors that we have for the measurements are a result of fitting nearly 10,000 data points from 28 detectors with a few parameters. The parameter $\sigma$ is a scale parameter in the final fit, and so does not have error bars associated with it. This parameter is typically uncertain to .5$\%$ if it is regarded as a fitting parameter. 

 \begin{figure}[p] 
   \centering
   \includegraphics[width=5in]{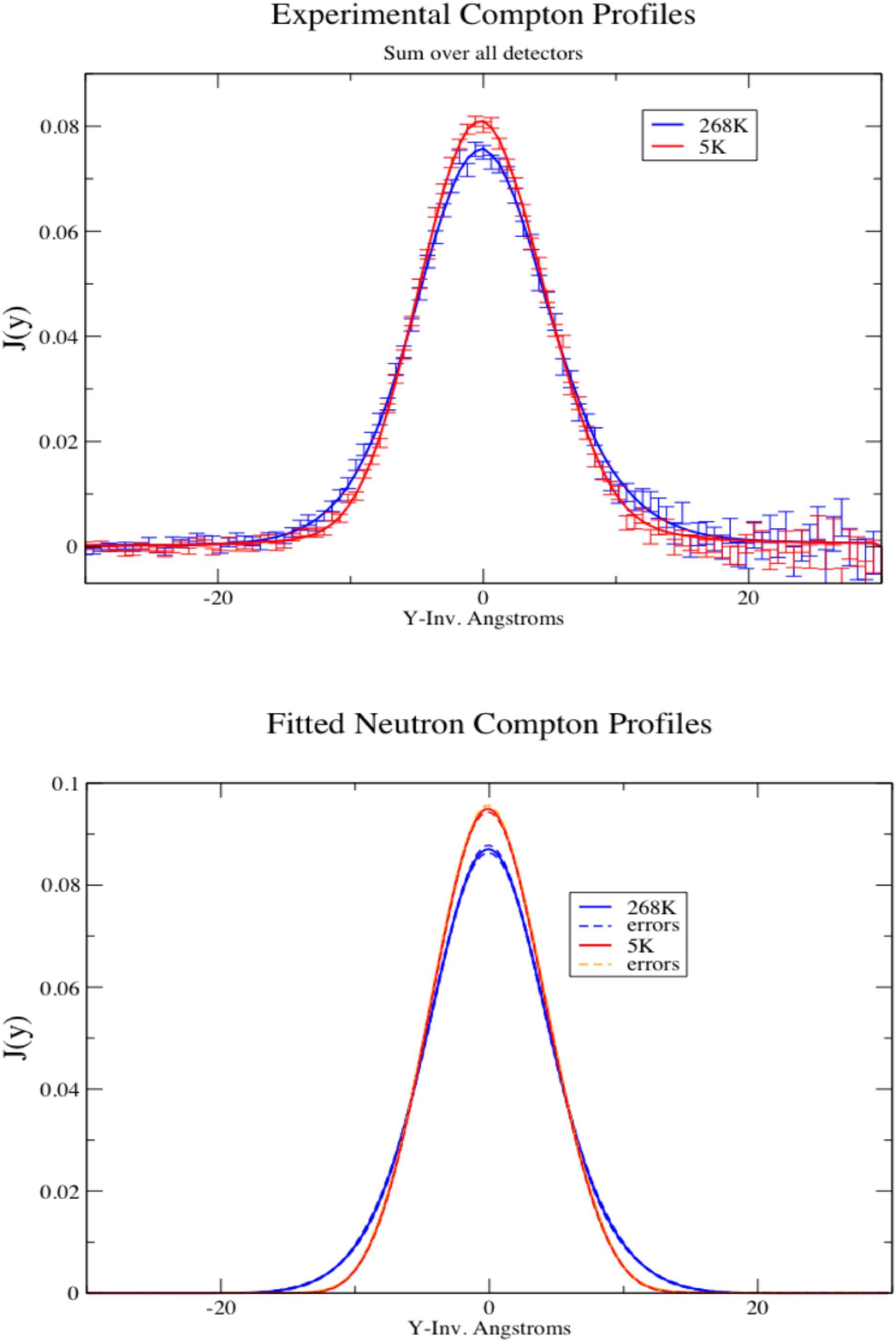} 
   \caption{(Top)The experimental Compton profile, J(y) at 5K and 268K, constructed by binning the time of flight data for each of the 28 detectors  in $\Delta y$ bins assuming that the data  satisfies y scaling(Eq. \ref{ysc}). The solid curves are fits to the entire data set, and include the instrumental resolution and final state effects. (Bottom) The extracted Compton profiles at 5K and 268K after fitting the entire data set. The resolution and the final state corrections that are contained in the fits to the experimental data  have been eliminated. The fits to the n(p) that results from the J(y) are described by Eq. \ref{inv} and the parameters for the fits are given  in Table 1. The errors shown are 1 standard deviation. }
   \label{fig:ice}
\end{figure}

\begin{table}
\small
\caption{Parameters for Fit}
\begin{tabular}{||c|c|c|c|c|c||}
Water Sample& $\sigma (\AA^{-1})$ &$a_2$& $a_3$&$a_4$&$a_{5}$ \\
\hline
Nanotube 5K&4.11&$-.053\pm .013$&$.041\pm .008$&0.0&$-.053\pm.03$\\
\hline
    Ice Ih 5K&4.79&$.018\pm .002$&$-.028\pm.003$&$.035\pm.004$&$-.037\pm.004$\\
\hline
HDA Ice 8K&4.74&0.0&0.0&$.051\pm.006$&$-.054\pm.007$\\
\hline
Ice VI 5K &4.66&$-.019\pm.005 $& $-.0067\pm.003 $&$.035\pm.004$&$-.057\pm.009$\\
\hline
Nanotube 268K & 4.82 &$.067\pm.010 $& $-.078\pm.012 $&$0.0$&$0.0$\\
\end{tabular}
\end{table}

We show in Fig. 2 a comparison of the fitted radial momentum distributions(4$\pi p^2 n(p)$) for the water in the   nanotubes at 5K with bulk polycrystalline ice Ih, ice-VI, and hda ice, all at comparable temperatures. While there are significant differences between the various forms of ice, the water in the nanotubes, presumably also a form of ice at 5K, is in a state that is qualitatively different from that of the the other forms of ice.  This shows up dramatically in the narrowness of the momentum distribution.   We find from Table 1 that the kinetic energy of a proton  in the nanotube-water is 35meV less than that  in ice Ih, and 44meV less than that in  the 268K phase.
 \begin{figure}[h] 
   \centering
   \includegraphics[width=5in]{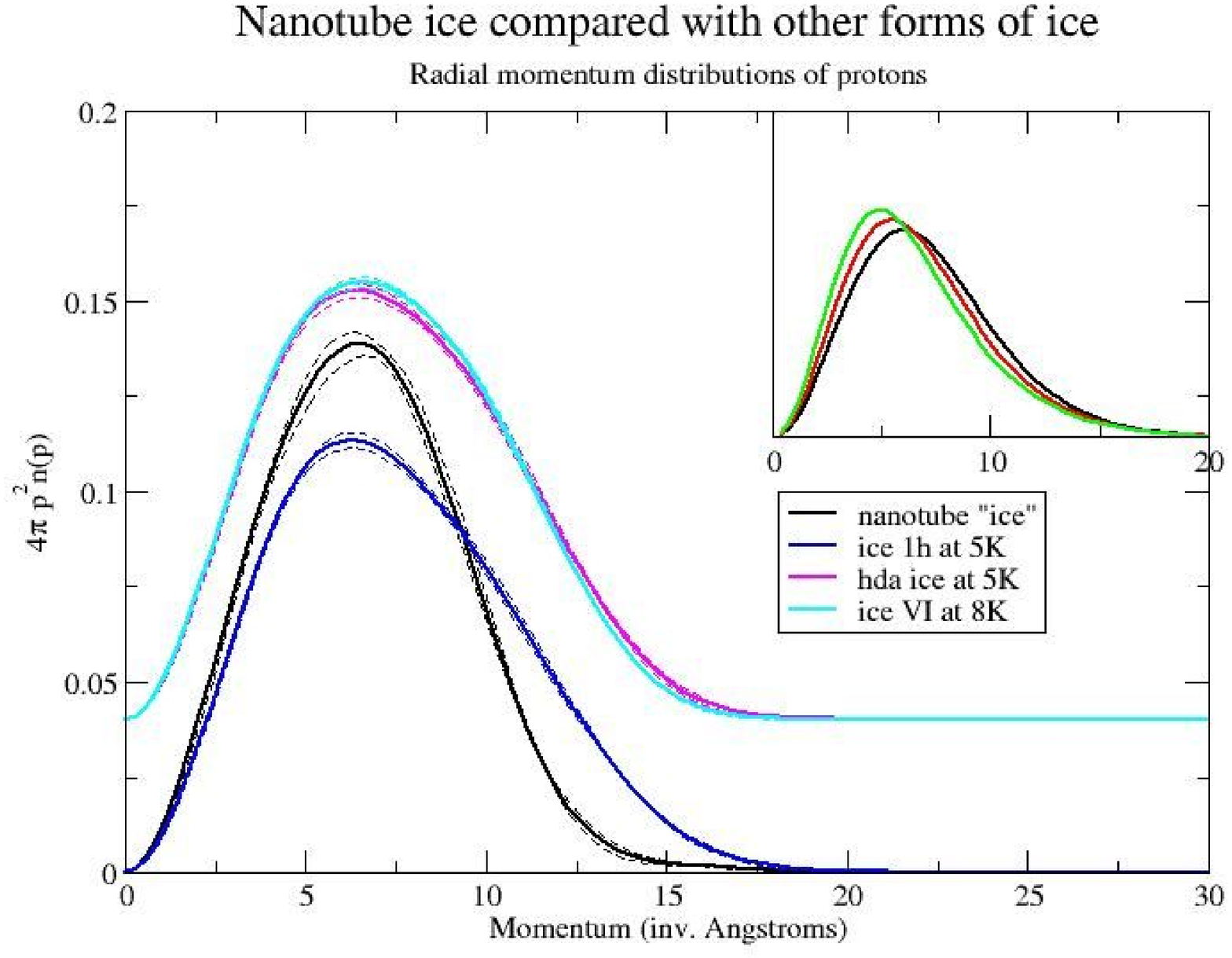} 
   \caption{The momentum distribution of the protons in nanotube ice compared with that in three forms of bulk ice. The dotted lines are one standard deviation error limits. The inset shows the effect of varying the parameters in an anisotropic  harmonic model of the proton momentum distribution in the hydrogen bond, with the momentum width along the bond fixed and the transverse widths varied.   The hda ice and ice VI data are displaced vertically for clarity. }
   \label{fig:ice}
\end{figure}

 Another significant difference, which is indicative of a major change in the local structure, is the absence of a  broad high momentum tail in the nanotube data.  If the motion along the O-H$\cdot\cdot$O bond and perpendicular to it was approximately harmonic, as  in ice Ih\cite{braz} at 269K,  the momentum distribution  for a single molecule would be described as  an anisotropic gaussian, the frequencies of motion perpendicular and parallel to the bond determining the momentum widths in the three directions. Since the stretch mode due to motion along the bond  is at a much higher frequency than the transverse modes(bending, libration), the momentum distribution will be much broader in the bond direction. When the individual molecule distribution is spherically averaged,  the shape of the resulting curve at momentum high compared to the transverse widths is determined entirely by the motion along the bond. This is shown in detail in the inset in Fig 2, where an isotropically averaged momentum distribution is shown for several values of the transverse momentum widths(2.8(green), 3.2(red) and 3.6 $\AA^{-1}$(black)) keeping the width along the bond of  6   $\AA^{-1}$.  The lowest value of the transverse width gives a kinetic energy comparable to that observed in the nanotubes ar 5K, the highest value corresponds to the fit to the ice Ih data.\cite{braz} 
 
 Evidently, an  harmonic model for the momentum distribution will always show a high momentum  tail due to the stretch mode. The forms of bulk ice are similar at high momenta because the covalent bond of the proton to its molecular oxygen is similar, and this is the dominant interaction determining the stretch mode frequency.   The absence of a high momentum tail, or rather its large reduction in intensity in the nanotube ice, indicates that the local structure of the water in nanotube ice is very different from that of  the other forms of ice. A best fit with an anisotropic gaussian distribution gives the widths along the bond and perpendicular to it as nearly equal and approximately 4 $\AA^{-1}$ The strong covalent molecular bond,  reponsible for the high frequency of the stretch mode, with a momentum width of approximately 6 $\AA^{-1}$, appears to be missing!   

\begin{figure}[h] 
   \centering
   \includegraphics[width=5in]{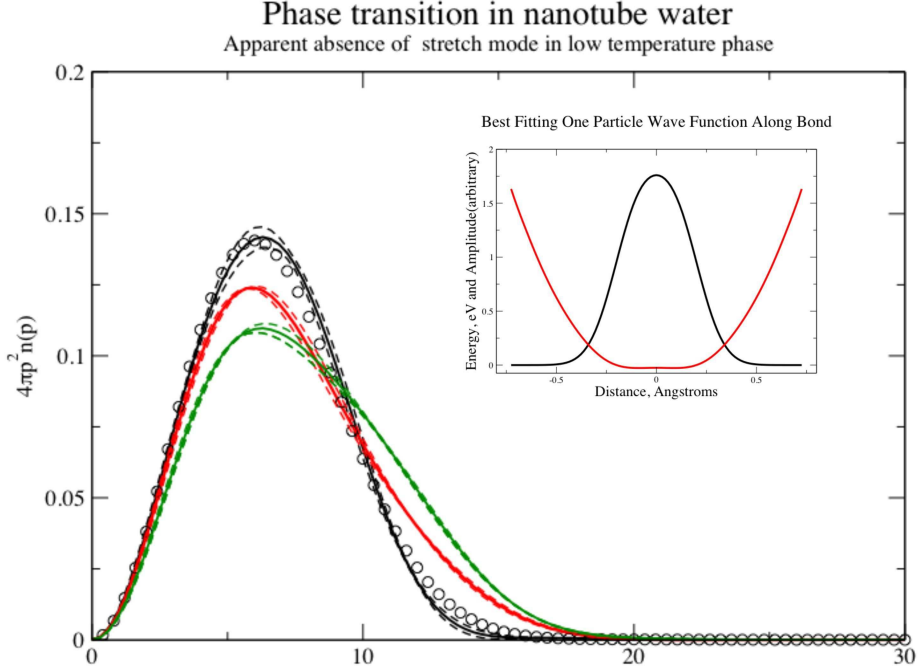} 
   \caption{ Comparison of the momentum distributions for nanotube ice at 268K(green), bulk ice Ih at 269K(red),  and the low temperature phase of nanotube ice, represented by the data at 5K(black). The circles are a fit to a model in which the proton is delocalised along the bond in a double well potential. The potential(red) and wave function(black) corresponding to the fit to the momentum distribution data along the bond direction are shown in the inset. }  
\label{fig:nano2}
\end{figure}
 
According to simulations, the initial water molecules entering the nanotube form a 2-D  square ice layer parallel to the walls of the nanotube.\cite{kog, no, ref1}.  The  carbon nanotubes used here  are sufficiently large(14$ \AA$) that a chain of water molecules can fit down the center, in addition to the ice layer.\cite{ref 1}  The absence of a high momentum tail in the momentum distribution, must be a property of the ice sheath, perhaps in conjunction with the central molecules,  since the sheath  constitutes roughly 85$\%$ of the molecules.  
 
 Measurements at higher temperatures provide further evidence  that the anomalous quantum state is associated with the 2-D ice layer. We have made DINS measurements of nanotube-water at 170K and 230K as well. These are indistinguishable, within the experimental uncertainty,  from those at 5K. However, the distribution at 268K, shown in Fig. 3a, is dramatically different, and has the high momentum tail that we  associate with the stretch modes. Evidently, the local structure around the proton has changed.  The MD simulations also show a change in the global structure of the water in the nanotubes above 200K,  from a two dimensional ice sheath and chain to a three dimensionally hydrogen bonded structure that resembles bulk water/ice. The existence of the phase change is  in agreement with earlier work.\cite{kog, man}. We show in Fig. 4a the change in the structure that the MD simulations\cite{ref1} predict.  The ice sheath is no longer present at 268K, and  the structure is approximately four fold coordinated, as in ordinary ice, This change is demonstrated more quantitatively in  Fig. 4b where the change with temperature of the radial distribution of the oxygen and hydrogen is shown. The broadening of the radial density distribution associated with the breakup of the ice layer is clearly visible. The structural transition occurring above 200K to a 3-D structure is evidently responsible for the qualitative change in the momentum distribution.  
  \begin{figure}[h] 
   \centering
   \includegraphics[width=5in]{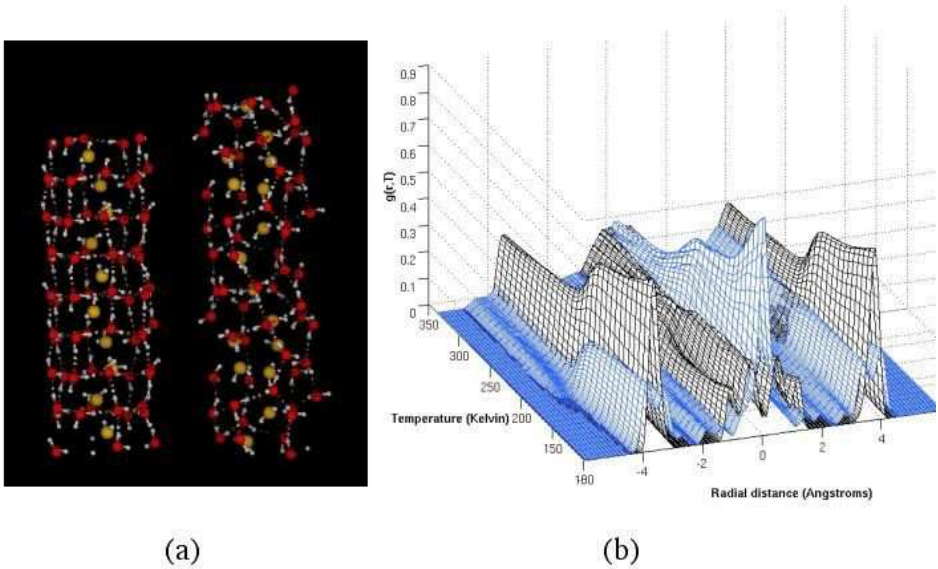} 
   \caption{(a)The simulated configuration of the water molecules in a smooth nanotube at 200K(left) and 268K(right)   The oxygen is red  (yellow if it is closer than 2.5$\AA$ to the axis), the hydrogen is white. Hydrogen bonds are dashed white lines. (b) The radial density distribution of the oxygen(dark brown) and hydrogen(light blue) as a function of temperature. The phase transition is visible as the knee in the graphs at which the proton distribution broadens}
   \label{fig:inset2}
\end{figure}

  We show also in Fig 3a, a fit to the data(circles) obtained from a simple model in which the wavefunction along the bond direction is the sum of two Gaussians displaced a distance d.  The distribution perpendicular to the bond is taken to be a Gaussian, with the two directions equivalent. The potential along the bond that would lead to such a wavefunction, and the wave function itself, are shown inFig 3b. The fitted values for the separation d is .21 Angstroms.  $\sigma_z$, the momentum width in the absence of any separation,  is 5.81$\AA^{-1}$, and the transverse momentum distribution  widths are 4.16 $\AA^{-1}$
  
It is not clear what is responsible for the change in the shape of the local potential for the protons, which is very nearly harmonic in water or ice.  The O-O separation  on average in the ice layer is 2.92$\AA$, estimated from the measured O-H stretching mode frequency\cite{ref1}, and is greater than that in ordinary ice.  If the simulations of the structure and this estimate are correct,  it is not due to very strong hydrogen bonds, as in ice X for instance.  Path integral Monte-Carlo calculations using the TTM2-F empirical potential for the water molecules and their interaction does not reproduce the momentum distribution data, as the strong covalent bond present in the isolated water molecule is only slightly modified by the interactions between molecules, and hence the momentum distribution has a high momentum tail\cite{cg}. This would be true of any other empirical potential that we know of as well. It is conceivable that the close proximity to the carbon nanotube wall is responsible for a readjustment of the electronic structure in the water molecules. It should also be remembered that the potential shown in Fig 3b is only an effective one particle potential.  The delocalization that we are seeing in the wavefunction and the flat bottom of the well could be the result of coherent interactions between the protons, perhaps involving the central chain molecules, which do form temporary weak H-bonds with the ice layer. Whatever the explanation, it is clear that the quantum state of the protons in the low temperature phase of water in these nanotubes is qualitatively different from that of any phase of water seen so far. The transition temperature to a normal bulk water/ice-like phase is likely to be dependent on the size of the nanotube and the details of the interaction of the water molecules with the confining cylinder. \cite{no}.  Should the phase exist at room temperatures in different size cylinders, its properties would be of great interest in understanding the structure and transport of water in biological pores.   
\centerline{\bf Acknowledgements:}
The work at Argonne National Laboratory was supported by the Office of Basic Energy Sciences, Division of Materials Sciences, U.S. Department of Energy, under Contract No. W-31-109-ENG-38. CB is thankful for Grant from Argonne Theory Institute for his stay at IPNS. The MD calculations were made with a grant of time on the Argonne National Laboratory Jazz computer. The work of G. Reiter, C. Burnham, D. Homouz and P. Platzman was supported by  DOE Grant DE-FG02-03ER46078

\end{document}